\documentclass[12pt]{iopart}
\usepackage{iopams,color}
\usepackage[dvips]{psfrag,graphicx}

\usepackage{verbatim}
\usepackage{subfigure}
\usepackage{hyperref}
\begin{document}
\title{Non-adiabatically driven electron in quantum wire with spin-orbit interaction}

\author{T \v Cade\v z$^1$ and J H Jefferson$^2$ and A Ram\v sak$^{1,3}$}

\address{$^1$ J. Stefan Institute, Ljubljana, Slovenia}
\address{$^2$ Department of Physics, Lancaster University, Lancaster LA1 4YB, UK}
\address{$^3$ Faculty of Mathematics and Physics, University of Ljubljana, Ljubljana, Slovenia}
\ead{tilen.cadez@ijs.si}

\begin{abstract}
An exact solution is derived for the wave function of an electron in a semiconductor quantum wire with spin-orbit interaction and driven by external time dependent harmonic confining potential.  The formalism allows analytical expressions for various quantities to be derived, such as spin and pseudo-spin rotations, energy and occupation probabilities for excited states. It is demonstrated how perfect spin and pseudo-spin f\mbox{}lips can be achieved at high frequencies of order $\omega$, the confining potential level spacing. By an appropriately chosen driving term, spin manipulation can be exactly performed far into the non-adiabatic regime. Implications for spin-polarised emission and spin-dependent transport are also discussed.
\end{abstract}

\pacs{71.70.Ej, 73.63.Kv}
% 71.70.Ej	Spin-orbit coupling, Zeeman and Stark splitting, Jahn-Teller effect
% 73.63.Kv	Quantum dots
\maketitle

One of the first proposals for the solid state realization of quantum information processing  and computation  is based on the use of the spin degree of freedom of electrons, confined in quantum dots (QD), which realize the qubit~\cite{Loss98}, and several universal quantum computation propositions~\cite{Levy02, DiVincenzo00, Stepanenko04} have been deduced from this basic idea.

Experimental progress in measurements and coherent manipulation  of electrons in QD systems has been immense in the past years.\cite{Koppens06, Nowack07, Nadj-Perge10, Berezovsky08, Press08, Brunner11, Hermelin11, McNeil11, Kataoka09, Kataoka11, Petta05, Hanson07, Bennett10} Coherent single electron spin rotations have been realized using oscillating magnetic~\cite{Koppens06} and oscillating electric fields.\cite{Nowack07, Nadj-Perge10} The limitation of these schemes is due to long single spin rotation times, of the order of tens of ns. By exploiting ultrafast laser control and optical pumping, single spin control was demonstrated on a ps timescale.\cite{Berezovsky08, Press08} All-electrical two-qubit gates composed of single-spin rotations and interdot exchange in a double QD were also demonstrated.\cite{Brunner11} By using surface acoustic waves, single electrons have been transfered between quantum dots separated by a few $\mu$m \cite{Hermelin11, McNeil11}, where McNeal {\it et. al.} have reliably relocated a single electron back and forth between the dots multiple times up to a cumulative distance of 0.25 mm.\cite{McNeil11} Controlled production of entangled qubit pairs in quantum dots and surface acoustic waves was predicted theoretically.\cite{Giavaras06, Jefferson06, Rejec00} Coherent time evolution of single electron charge dynamics on a time scale of a few picoseconds was measured.\cite{Kataoka09} Tunable non-adiabatic single electron excitations were observed recently,\cite{Kataoka11} where rapid modulation of confining potential causes transitions to excited orbital states.

Due to the lack of inversion symmetry in semiconducting materials, the electron spin is coupled to the orbital motion.\cite{Bychkov84, Dresselhaus55} Rather than treating this interaction as a source of decoherence, it can be harnessed to coherently manipulate spin-orbital qubits.\cite{Stepanenko04, Nadj-Perge10, Flindt06, Coish06, San-Jose08, Golovach10, Bednarek08,Nadj-Perge12} The rotation of a spin-orbital qubit in a moving quantum dot has been studied analytically in the adiabatic limit. Single qubit manipulation, spin-flip for example, can easily be performed by adiabatic spatial QD translation for a distance of the order of the spin-orbit length.\cite{Coish06, San-Jose08, Golovach10, Bednarek08} However, the condition of adiabaticity sets a severe upper limit for the speed of single qubit operations. Although the analysis by means of a full numerical simulation is possible, it gives results only for particular cases and the optimization and elimination of errors is not straightforward.\cite{Bednarek08}

\begin{figure}[bht]
\centering
\includegraphics[width=0.75\textwidth]{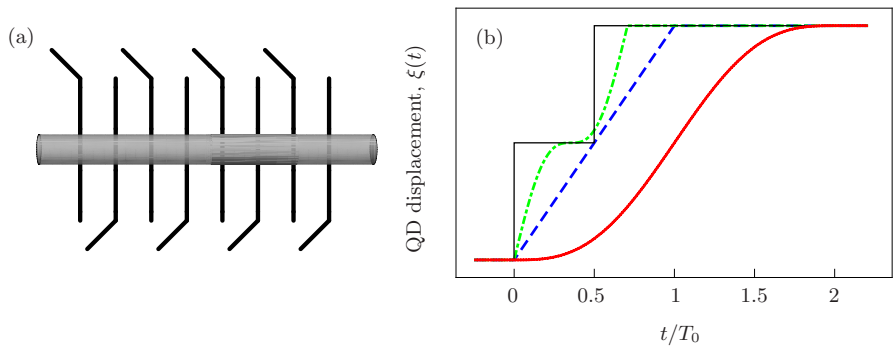}
\caption{(a): Schematic view of  a semiconductor quantum wire deposited on gate electrodes which provide a time dependent harmonic potential. Such systems have been realized experimentally~\cite{Nadj-Perge10, Nadj-Perge12, Fasth05}. (b): Position $\xi(t)$ of the harmonic potential minimum.  After driving time $T$ the QD stops at displacement $\xi(T)$. Black line corresponds to the fastest unidirectional QD translation, with minimum spin-flip time $T_0/2$, where $T_0 = 2 \, \pi /\omega$ and $\omega$ is the confining potential level spacing. Besides instantaneous displacement, the other drivings considered are: linear ramp  with $\xi(t)=\xi(T) t/T$ (dashed blue) and 
sinusoidal with $\xi(t) = \xi(T)\, \bigl[ t/T \pm \, \sin(2 \, \pi \, t/T)/(2 \, \pi) \bigr]$ (dot-dashed green and solid red), giving spin-flip times of $T_0$, $T_0/\sqrt{2}$ and $2 \, T_0$, respectively.}\label{fig:wire}
\end{figure}

In this paper we extend analytic results for spin manipulation by adiabatic translation of a QD\cite{Coish06, San-Jose08, Golovach10} to include the non-adiabatic regime, thereby opening the possibility of much faster spin-qubit rotations. In particular, we analytically solve the one-dimensional (1D), non-adiabatically driven QD with harmonic confining potential in the presence of spin-orbit interaction. Although it is demonstrated how a general solution can be constructed, we also present explicitly four typical driving schemes with the focus on the non-adiabatic regime.
The system considered here is similar to the one presented by Flindt {\it et. al.},\cite{Flindt06} where the authors performed a spin-f\mbox{}lip by displacement of a QD in a wire. Such a system has even been realized experimentally,\cite{Nadj-Perge10, Nadj-Perge12, Fasth05, Fasth07, Shin12} with the ability to control a single electron, in InAs~\cite{Fasth07} and Ge~\cite{Shin12} quantum wires. More recently, spin-orbital qubits in double QD have also been demonstrated in InSb quantum wires.\cite{Nadj-Perge12}

The 1D Hamiltonian we consider is given by
\begin{equation}\label{Ht}
H(t) = \frac{p_{x}^2}{2 \, m^*} + \frac{m^* \, \omega^2}{2} \, \bigl[{x} - \xi(t) \bigr]^2 +  p_{x} \, \bigl(\alpha \, \sigma_{\mathrm y} - \beta \, \sigma_{x} \bigr),
\end{equation}
where an electron with an ef\mbox{}fective mass $m^*$ is confined in a harmonic trap with frequency $\omega$, centered around the time dependent position of harmonic potential minimum $\xi(t)$,
$\alpha$ and $\beta$ are the Rashba~\cite{Bychkov84} and Dresselhaus~\cite{Dresselhaus55} spin-orbit couplings, respectively, and we use units with $\hbar=1$ throughout the paper.

First we apply the unitary transformation 
\begin{equation}\label{U^+}
U^{\dagger}(t) = e^{- S} \, V^{\dagger}(t),
\end{equation}
which consists of a canonical transformation that eliminates the spin-orbital part of the Hamiltonian,\cite{Khaetskii00}
\begin{equation}\label{S}
S = i \, m^* \, {x} \, (\alpha \, \sigma_{\mathrm y} - \beta \, \sigma_{x}),
\end{equation}
and of the part which eliminates the time dependent part of the Hamiltonian,\cite{teerHaar} 
\begin{eqnarray}
V^{\dagger}(t) & = & e^{i \, \Phi (t)} \, e^{i \, m^* \, \dot{x}_c (t)\, x} \, e^{-i \, x_c(t) \, p_x}, \nonumber \\
\Phi (t) & = & - \int_0^t \, L(t') \, \mathrm{d}t', \\
L(t) & = & \frac{m^*}{2} \,  \dot{x}_c(t)^2 - \frac{m^* \, \omega^2}{2} \, \left[ {x}_c(t)^2 -  \xi(t)^2\right]. \nonumber
\end{eqnarray}
Here $x_c(t)$ is the solution of the  equation of motion for a classical particle in a driven harmonic oscillator potential,
\begin{equation}
\ddot{x}_c + \omega^2 \, x_c =\omega^2 \, \xi(t), \label{classical}
\end{equation}
for $x_c(0)=0$ and $\dot{x}_c(0)=0$ customarily given by $x_c(t)=\omega\int_0^t \sin[\omega(t-t')] \, \xi(t') \, \mathrm{d}t'$.

The time dependent Schr\" odinger equation after the unitary transformation $U(t) H(t) U^{\dagger}(t) \to H$ takes the form of a static harmonic oscillator 
\begin{equation} \label{hpsi}
i \, \partial_t \, \psi = \Bigl[ \frac{p_{x}^2}{2 \, m^*} + \frac{m^* \, \omega^2}{2} \, x^2 +E_{so} \Bigr] \, \psi,
\end{equation}
with an energy shift $E_{so}= - m^* \, (\alpha^2 + \beta^2)/2$. The exact eigenstates of the original Hamiltonian \eref{Ht}, $\Psi_{n s}(x, t)$, are obtained directly from the eigenstates of \eref{hpsi}, $\psi_n(x,t)$, via the unitary transformation \eref{U^+}, {\it i.e.},
\begin{eqnarray}\label{wavefunction}
\Psi_{n s}(x, t) & = & e^{-i \, (E_{so}+\omega_n) \, t} \, U^{\dagger}(t) \, \psi_n(x) \, \chi_s =\nonumber \\
& = & e^{i \, \varphi_n(x, t)} \, \psi_n[x - x_c(t)]  \, e^{-S} \, \chi_s,
\end{eqnarray}
where $\varphi_n(x, t) =-(E_{so} + \omega_n) t + \Phi (t) - m^* \dot{x}_c(t) \, x$ is a phase factor,  $\omega_n =  (n + 1/2)\omega$ is the eigenenergy of the $n$-th eigenfunction of the static harmonic oscillator $\psi_n$, and $\chi_s$ a spinor with spin $s$. 

From the exact time-dependent solution, \eref{wavefunction}, we see that the effect of non-adiabatic motion is to displace the harmonic oscillator ground-state gaussian wavepacket away from the quantum-dot minimum at $\xi(t)$ to a new position $x_c(t)$, given by the solution of the classical oscillator, \eref{classical}. It is clear that this will excite the electron in the QD and explicit calculation gives
\begin{eqnarray}
{\cal E}_n(t) & = & \bigl< \Psi_{n s} \bigl| H(t) \bigr| \Psi_{n s} \bigr> =  \nonumber \\ & =&E_{so} + \omega_n +  \frac{m^* \, \dot{x}_c(t)^2}{2} + \frac{m^* \, \omega^2}{2} \, \bigl[ x_c(t) - \xi(t) \bigr]^2,\label{e}
\end{eqnarray}
which is in fact the classical expression for the harmonic oscillator energy, apart from a constant shift of $E_{so} + \omega_n$. To simplify analysis we consider quantum wires grown in the $[111]$ direction, for which $\beta = 0$. For this case the mean spin rotates around the $y$-axis by an angle $2 \, x_c(t)/\lambda_{so}$, where $\lambda_{so} = (m^* \, \alpha)^{-1}$, following immediately from equation \eref{wavefunction}, from which the expectation values of the spin components may also be evaluated explicitly. For example, if the initial state is the pseudo-spin-up Kramers state, $\Psi_{0 \uparrow}(x, 0) = e^{- i \, x/\lambda_{so} \, \sigma_y} \, \psi_0 (x) \, \chi_{\uparrow}$, the spin expectation values are 
\begin{eqnarray}\label{spin}
{\cal S}_z(t)=\frac{1}{2}\langle \sigma_z \rangle & = & \frac{1}{2}\cos\bigl[2 \, x_c(t)/\lambda_{so} \bigr] \, e^{-2 \, \sigma^2/\lambda_{so}^2}, \nonumber \\
{\cal S}_x(t)=\frac{1}{2}\langle \sigma_x \rangle & = & \frac{1}{2}\sin\bigl[2 \, x_c(t)/\lambda_{so} \bigr] \, e^{-2 \, \sigma^2/\lambda_{so}^2},
\end{eqnarray}
where the oscillatory factors are consistent with the mean spin rotation and the decaying exponential factors account for the spread in the wavefunction, which is given by $\sigma = (m^* \, \omega)^{-1/2}$. Due to the precession of spin around the $y$-axis ${\cal S}_y$ is constant and zero for our choice of the initial state.

In figure~\ref{fig:wire}(b) we show specific choices of driving term, $\xi(t)$, for cases when the transit time, $T$, is chosen to give a perfect spin-flip, for which the dot displacement is $\pi\lambda_{so}/2$. Detailed results for these choices and general $T$ are given below.
Here we note that in general there will be residual oscillations of the electron distribution after the dot stops, leading to uncertainty of order $2 \, a/\lambda_{so}$ in the rotated spin rotation angle , where $a$ is the amplitude of the residual orbital oscillations. Such residual oscillations of the electron wave packet after the dot has stopped will also give rise to orbital decoherence through phonon emission and subsequent spin decoherence via the spin-orbit interaction. However, it is clear that $\xi(t)$ can always be chosen such that these residual oscillations are totally suppressed with the wavefunction returning to the centre of the QD and remaining there, {\it i.e.}, $x_c(t)=\xi(T)$ and $\dot{x}_c(t) = 0$ for all $t > T$. This is a very physical result which can be understood in purely classical terms from equation \eref{classical}. The final energy is given by ${\cal E}_{\mathrm{final}} = E_{so} + \omega_n +m^* \, \omega^2 \, a^2/2$.

A further interesting general observation is that under forced oscillations in which the quantum dot is translated back and forth non-adiabatically in one dimension, energy will be transferred to the electron wave packet in the dot and under resonant conditions the amplitude of the displacement of the classical oscillator (and hence the position of the wavefunction relative to the potential minimum) will increase, as will the mean energy of the electron given by equation \eref{e}.  Accompanying these oscillations will be a corresponding rotation of the spin which will also have oscillatory components, due to the oscillation of the dot and oscillation of the wavefunction relative to the dot. It would be interesting to observe effects of these oscillations experimentally, {\it e.g.} in the polarisation distribution of radiative emission when the excited electron in the quantum dot relaxes or by using weak repeated measurement of current.~\cite{Kataoka09, Kataoka11} 

In materials with spin-orbit interaction, besides the real spin, one can study also the pseudo-spin, defined by
\begin{eqnarray}
{\cal \widehat{T}}_z& = &\frac{1}{2} \, \sum_{n=0}^\infty \, (|\widetilde\Psi_{n\uparrow}\rangle\langle\widetilde\Psi_{n\uparrow}|-|\widetilde\Psi_{n\downarrow}\rangle\langle\widetilde\Psi_{n\downarrow}|), \nonumber \\ 
{\cal \widehat{T}}_x&=&\frac{1}{2} \, \sum_{n=0}^\infty \, (|\widetilde\Psi_{n\uparrow}\rangle\langle\widetilde\Psi_{n\downarrow}|+|\widetilde\Psi_{n\downarrow}\rangle\langle\widetilde\Psi_{n\uparrow}|),\label{oppseudo}
\end{eqnarray}
where $\widetilde\Psi_{n s}(x,t)$ are the eigenstates of the potential at a particular instant $t$,
\begin{equation}
\widetilde\Psi_{n s}(x,t) = e^{- i \, m^* \, \alpha \, [x - \xi(t)] \, \sigma_y} \, \psi_n\bigl[ x - \xi(t) \bigr] \, \chi_s.\label{basis}
\end{equation}
For the initial Kramers pseudo-spin-up state, a straightforward calculation leads to a simple exact expression for the expectation value,

\begin{eqnarray}
{\cal T}_z(t) = \langle {\cal \widehat{T}}_z \rangle & = & \frac{1}{2}\cos[2 \, \xi(t)/\lambda_{so}], \nonumber \\
{\cal T}_x(t) = \langle {\cal \widehat{T}}_x \rangle & = & \frac{1}{2}\sin[2 \, \xi(t)/\lambda_{so}],\label{pseudospin1}
\end{eqnarray}
and ${\cal T}_y=0$. Pseudo-spin is dependent only on the position of the QD, therefore slow adiabatic or arbitrary fast non-adiabatic motion lead to identical pseudo-spin values. Furthermore, the pseudo-spin is independent of the spread in orbital wavefunction, being a composite of spin and orbital degrees of freedom. In the derivation of this result we express the coefficients in the instantaneous basis given by \eref{basis},
\begin{eqnarray}
c_{n s}(t) & = & \bigl< \widetilde\Psi_{n s} \bigr| \Psi \bigr> = e^{- i \, [(E_{so} \, + \, \omega/2) \, t - \Phi(t)]} \, \chi_s^{\dagger} \, e^{- i \, \xi(t)/\lambda_{so} \, \sigma_y} \, \chi_{\uparrow} \, I_n(t), \label{coeff} \\
I_n(t) & = & \int \, \mathrm{d}x \, e^{- i \, m^* \, \dot{x}_c(t) \, x} \, \psi_n^*\bigl[ x - \xi(t) \bigr] \, \psi_0[x - x_c(t)] = \nonumber \\
& = & \, e^{- \dot{x}_c(t)^2 /(2 \, \sigma^2 \, \omega^2)+{i \, m^* \, \dot{x}_c(t) \, x_c(t)}} \, \langle n \bigl| 0 \rangle_d, \nonumber
\end{eqnarray}
where $\langle n \bigl| 0 \rangle_d= \int \, \mathrm{d}u \, \psi_n^*(u)\, \psi_0(u-d)$ is the overlap of the $n$-th eigenfunction of the quantum harmonic oscillator and the ground state displaced by the distance $d =  x_c(t)-\xi(t)  - i \, \dot{x}_c(t)/\omega$, and $\langle n \bigl| 0 \rangle_d  = [d/(\sqrt{2} \sigma)]^n/\sqrt{n!} \, \exp[- d^2/(4 \, \sigma^2)]$. Note that this result follows due to the independence of the spin and the orbital part. Thus the probability of occupation of the $n$-th manifold is exactly determined solely by classical coordinates, $P_n(t)=\sum_s |c_{n s}(t)|^2=|I_n(t)|^2$.

From equation \eref{coeff} we see that the exact wavefunction at time $t$ may be written,
\begin{eqnarray}\label{instantaneous_basis}
\Psi(x,t) & = & \sum_{n s} \, c_{n s}(t) \, \widetilde\Psi_{n s}(x,t) = \nonumber \\
& = & e^{- i \, [(E_{so} + \omega/2) \, t - \Phi(t)]}\, \bigl[ \cos(\xi(t)/\lambda_{so}) \, \widetilde\Psi_{\uparrow} + \sin(\xi(t)/\lambda_{so}) \, \widetilde\Psi_{\downarrow} \bigr],
\end{eqnarray}
where $\widetilde\Psi_{s} = \sum_{n} \, I_n(t) \, \widetilde\Psi_{n s}(x,t)$. In this form we see explicitly the rotation of the pseudo-spin on the Bloch sphere and also the time-dependence of the orbital part of pseudo-spin states $\widetilde\Psi_{s}$.

Pseudo-spin qubits in systems with strong spin-orbit interaction are usually restricted to the manifold spanned by the ground state Kramers doublet, as studied in references~\cite{Nadj-Perge10, Flindt06, Coish06, San-Jose08, Golovach10, Bednarek08}. Pseudo-spin \eref{oppseudo} is in this case limited only to the $n=0$ manifold. For the initial Kramers pseudo-spin-up ground state $\Psi_{0\uparrow}(x,0)$, the expectation value of pseudo-spin is related to the probabilities $P_{0\uparrow} = \bigl| c_{0 \uparrow} \bigr|^2$ and $P_{0\downarrow} =  \bigl| c_{0 \downarrow} \bigr|^2$ that the system is in the ground state with pseudo-spin up or down, respectively. 

Direct evaluation gives 
${\cal T}_z^0(t) = \frac{1}{2} (P_{0\uparrow}-P_{0\downarrow})$, where  $P_{0 \uparrow} = \cos^2(\xi/\lambda_{so}) \, P_0$, $P_{0 \downarrow} = \sin^2(\xi/\lambda_{so}) \, P_0$, which leads to
\begin{equation}\label{pseudospin}
{\cal T}_{i}^0(t) =  {\cal T}_{i}(t)P_0(t), \quad i=x,y,z,
\end{equation}
and the probability for the ground state manifold occupation is given by $P_0(t) ={\exp}[-(x_c-\xi)^2/(2 \, \sigma^2)-\dot{x}_c^2/(2 \, \sigma^2 \, \omega^2)]$. 
Note that in the adiabatic limit the moving electron remains at the potential minimum, {\it i.e.}, $x_c(t) \to \xi(t)$ and $P_0 \to 1$ and the pseudo-spin coincides -- up to an exponential factor due to a wavefunction spread -- with the real spin, given by~\eref{spin}.

To be specific, we examine four examples of the driving term $\xi(t)$: sudden displacement in two steps, a constant velocity of the QD displacement, and two different types of smooth driving with sinusoidal (AC) type of velocity, as shown in figure~\ref{fig:wire}(b). In all cases the QD stops moving after the transit time $T$. Then, the position expectation value oscillates with the amplitudes of residual oscillations $a$, as follows from the classical solution  $x_c(t)$ of the equation of the motion \eref{classical}. For example,
\begin{eqnarray}
a_c & = & \frac{\lambda_{so}}{2} \, \frac{T_0}{T} \, \biggl| \sin\Bigl(\pi \, \frac{T}{T_0}\Bigr) \biggr|, \nonumber \\
a_s & = & \frac{T_0^{2}}{T^2 - T_0^2} \, a_c,
\end{eqnarray}
for constant velocity ($a_c$) case and smooth sinusoidal case ($a_s$), as introduced in figure \ref{fig:wire}  (solid red).

\begin{figure}[hbt]
\centering
\includegraphics[width=1\textwidth]{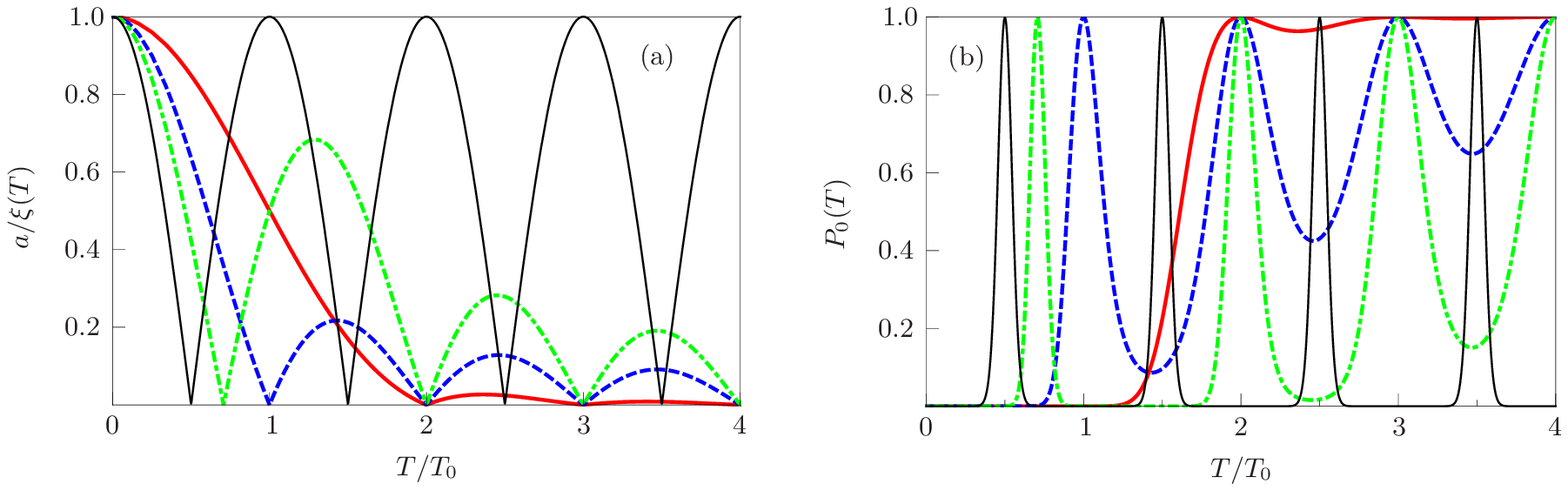}
\caption{(a): Amplitude of residual oscillations after the QD stops, $a=\sqrt{[x_c(T)-\xi(T)]^2+\dot{x}_c(T)^2/\omega^2}$, divided by the QD displacement $\xi(T) = \pi \, \lambda_{so}/2$ for drivings as in figure~1(b): Fastest unidirectional driving (full black), constant velocity (full red) and the two sinusoidal drivings, a broken (dot-dashed green) and a smooth one (dashed blue). (b): Probability for the ground state doublet occupation after the QD stopes. 
The material parameters that we use are given by $m^* = 0.015 \, m_e$ and $\lambda_{so} = 150$ nm, where $m_e$ is electron mass, which are typical for InSb,~\cite{Nadj-Perge12}, the confining potential level spacing $\omega = 10$ meV and $\sigma = 0.1 \, \xi(T)$.}\label{fig:residual}
\end{figure}

The shortest time with no residual oscillations for unidirectional movement of the dot, $T_{\mathrm{min}}$, is achieved by instantaneous displacement of the dot for $\pi \, \lambda_{so}/4$, followed by a waiting time $T_0/2$, followed by another instantaneous displacement for $\pi \, \lambda_{so}/4$. For any other unidirectional driving the minimum time $T_{\mathrm{min}}$ rises. In the limit $T \to 0$ all unidirectional movements of the dot lead to oscillations with the amplitude $\xi(T)=\pi \, \lambda_{so}/2$. In figure~\ref{fig:residual}(a) residual oscillations are presented as a function of driving time $T$. In all cases the residual oscillations vanish when some resonant condition for $T/T_0$ is fulfilled. The oscillations are proportional to the spin-orbit length $\lambda_{so}$ and have a simple dependence on $T_0$ and $T$. As expected, smooth sinusoidal driving exhibits significantly less residual oscillations compared to other forced oscillations considered. As demonstrated here, by using the solution to a classical driven harmonic oscillator we gain insight into a driven QD in a semiconducting nanowire. 

\begin{figure}[hbt]
\centering
\includegraphics[width=0.65\textwidth]{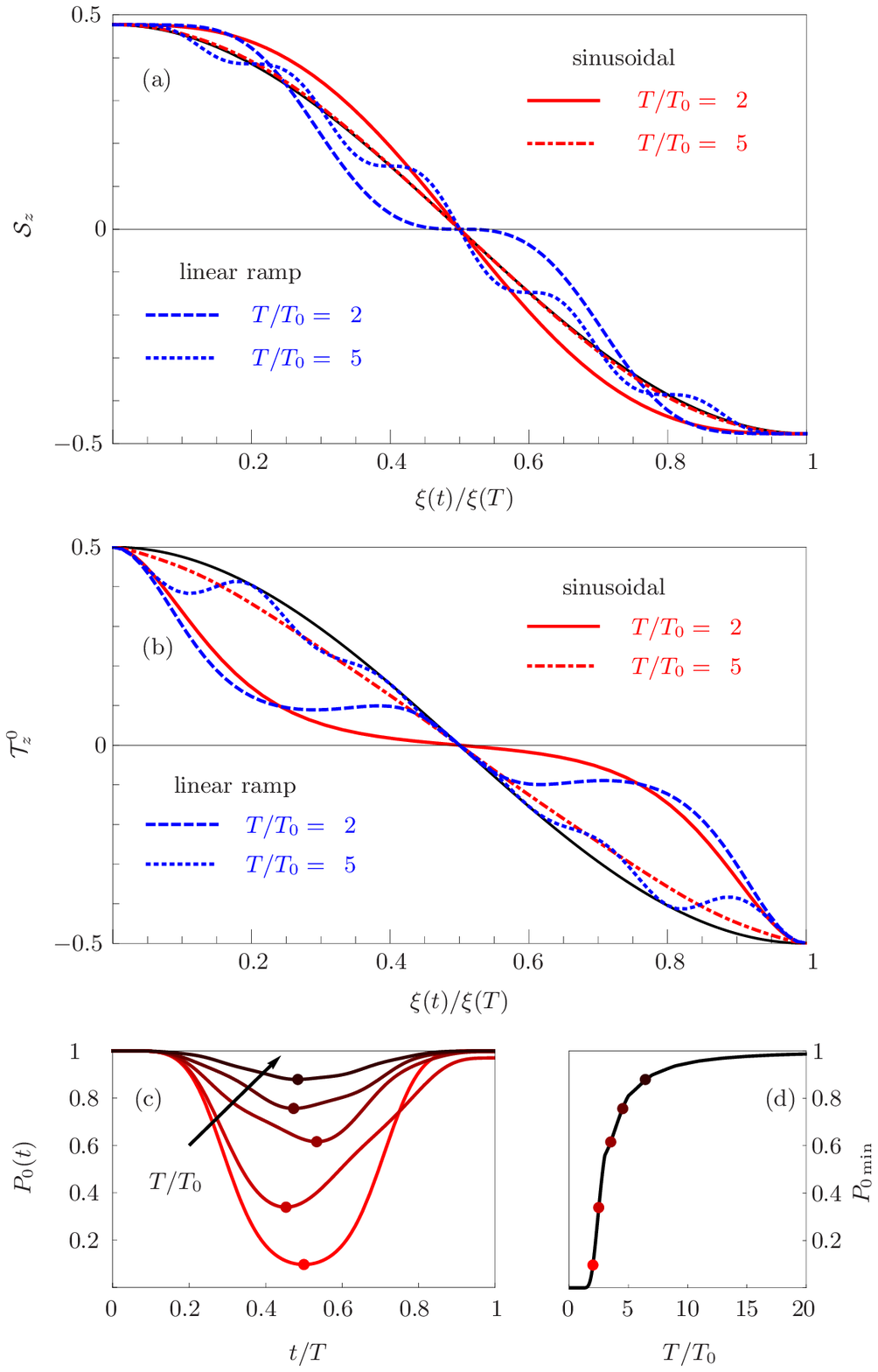}
\caption{(a) $z$-component of spin expectation value as a function of QD displacement in non-adiabatic regime for two types of driving. Full black line represents the adiabatic limit ($T\to\infty$). Note the oscillations superimposed on the cosine curve giving rise to both positive and negative fluctuations in mean spin relative to the adiabatic limit. (b) Ground-doublet pseudo-spin expectation, ${\cal T}_z^0$, for parameters as in (a).  Note that  ${\cal T}_z^0(0)=-{\cal T}_z^0(T)=\frac12$ precisely and that the effect of non-adiabaticity leads to greater deviation from cosine (compared with real spin) at intermediate times, which is {\it always} suppressed due to breakdown of the ground-doublet approximation. (c): Typical probability for occupation of the ground state doublet during the driving (sinusoidal), $P_0(t)$, in the non-adiabatic regime. The arrow indicates the direction of increasing $T$  ($T/T_0$=2, 2.5, 3.5, 4.5, 6.5); bullets indicate minimum values $P_{0 \, \mathrm{min}}$. (d): $P_{0 \, \mathrm{min}}$ as a function of $T/T_0$; bullets as in (c).  All parameters as in figure~\ref{fig:residual}(b). }\label{fig:adiabatic}
\end{figure}

In figure \ref{fig:adiabatic}(a) the $z$-component of spin after one cycle is presented for linear ramp and smooth sinusoidal driving for $T/T_0 = 2$ and 5. For this spin-flip case the quantum dot is displaced a distance $\xi(T)=\pi \, \lambda_{so}/2$. Note that the magnitude of ${\cal S}_z$ does not quite reach $\frac12$ at the beginning and end of the spin-flip due to the spread in the gaussian wavefunction, which gives ${\cal S}_z(0)=\frac12 \exp{(-2 \, \sigma^2/\lambda_{so}^2})$ from equation \eref{spin}.  Note also the weak oscillation superimposed on the cosine curve which, again, can be understood directly from the exact solution, equation \eref{spin}, which shows a perfect cosine variation when plotted against $x_c$ rather than $\xi$. For $x_c-\xi$ small we may expand to lowest order giving $\cos(2 x_c/\lambda_{so})=\cos(2 \xi/\lambda_{so}) -2 \, (x_c-\xi) \, \sin(2 \xi /\lambda_{so})/\lambda_{so}$ which is the form shown in the figure, since $x_c-\xi$ has $n=T/T_0$ oscillations. These oscillations are both positive and negative with respect to the adiabatic limit (full black line) which shows a perfect cosine (also with amplitude attenuated by the gaussian
factor in equation \eref{spin}) but with no superimposed oscillation since $\xi(t)=x_c(t)$ for all $t$.

In figure \ref{fig:adiabatic}(b) the ground-doublet pseudo-spin expectation ${\cal T}_z^0$ is presented for the same conditions as in (a). The adiabatic limit (full black line) is now a perfect cosine with amplitude $\frac12$ and this is also the result for full pseudo-spin ${\cal T}_z$ for any $T$ consistent with spin-flip (see equations \eref{oppseudo} and \eref{pseudospin1}). On the other hand, we see that there are significant deviations for non-adiabatic motion in the ground-doublet approximation, signaling its breakdown due to suppressed probability of ground-doublet occupation and corresponding suppression of ${\cal T}_z^0$ relative to the adiabatic limit.
These deviations do not occur, however, at the beginning and end of the spin-flip for which the function $\xi(t)$ have been chosen so that the electron dot starts and ends in its ground-doublet located in the centre of the dot.  Nevertheless, we see from the figure (\ref{fig:adiabatic}(a) and \ref{fig:adiabatic}(b)) that any error in the total displacement will give spin-flip errors which are significantly smaller for the smooth sinusoidal choice of $\xi(t)$, as expected.

For all the different driving functions $\xi(t)$, the optimum spin (or pseudo-spin) transformations, which leave the electron in its lowest energy state after transit of the dot, may always be determined.  Such states depend only on the distance the dot travels, independent of the degree of non-adiabaticity, provided the motion is coherent.  However, the  states of the electron in the dot at intermediate times/distances do depend strongly on the degree of non-adiabaticity and a convenient measure of this is the deviation from unity of the probability for the ground state, $P_0$. For $t\geq T$ this probability is simply determined by the residual oscillations, $P_0(T)=e^{-a^2/(2 \, \sigma^2)}$, as shown in figure~\ref{fig:residual}(b) for driving functions as  in figure~\ref{fig:residual}(a). In figure~\ref{fig:adiabatic}(c) the time dependence of $P_0(t)$ for different values of $T/T_0$ is presented for the smooth sinusoidal case: for large $T$, $P_0\sim1$, while for short times the minimum of $P_0=P_{0 \, \mathrm{min}}$ around $t\sim T/2$ signals the increase in probability for exciting higher energy states. The transition from highly non-adiabatic to the adiabatic regime is clearly seen also from the behaviour of the minimum value of $P_0 $, plotted in figure~\ref{fig:adiabatic}(d), which at high $T$ monotonically reaches the unitary limit.

To conclude, we have presented exact solutions of an electron in a driven QD in a 1D quantum wire with spin-orbit interaction. The solutions can be expressed analytically for a broad class of driving schemes. The essence of the full quantum mechanical solution is in the solution for the corresponding classical harmonic pendulum and by applying the appropriate canonical transformation. Exact analytical formulae for spin, pseudo-spin, average energy and probabilities for the occupation of excited states are given in terms of the coordinates of the QD and the position of the corresponding classical pendulum. 
To be specific, we mainly focused on two typical driving regimes: a constant velocity and a smooth sinusoidal driving -- in both cases in the full range from adiabatic to non-adiabatic driving. We demonstrated that perfect spin-flip can be realized even in the far non-adiabatic regime, providing driving which by the virtue of the formalism can easily be appropriately tuned. The solution given here may be used to guide more complicated simulations for the design of realistic gated device structures. 

Two extreme non-adiabatic regimes were highlighted. The first is when the forced oscillator is chosen such that after the transit time of the quantum dot the wavefunction of the electron is at the bottom of the QD potential well and remains there. The main advantage of this choice is that the associated spin rotation (single-qubit transformation) may be determined with maximum precision and does not subsequently change with time. The other extreme regime is exactly the opposite in which the forced oscillations have maximum amplitude after the transit time (resonance). Indeed, we can increase this amplitude indefinitely using a periodic driving term for the dot motion when it is resonant with the harmonic oscillations of the electron in the QD. The amplitude of these oscillations will, of course, be ultimately limited by dissipative processes, in particular phonon and radiative emission, though in clean systems coherent oscillations should survive to large amplitudes. What is special about such systems is the intimate connection between motion of the dot, the electron probability distribution relative to the dot and spin rotation. It would be interesting challenge to observe the effects of such charge-spin oscillations experimentally, for example in spin-polarised emission or spin-dependent transport.

\ack
The authors thank T Rejec for discussions and acknowledge support from the Slovenian Research Agency under contract P1-0044 and the EU grant NanoCTM.

\section*{References}

\bibliographystyle{iopart-num}
\bibliography{Cadez-bib}
\end{document}